\documentclass[journal,10pt]{IEEEtran}

\usepackage{graphicx}
\usepackage{cite}
\usepackage{amsmath}
\usepackage{amssymb}
\usepackage[square,sort&compress,comma]{natbib}
\usepackage{verbatim}
\usepackage{multirow}
\usepackage{diagbox}
\usepackage{booktabs}
\usepackage{color}

\usepackage[ruled]{algorithm2e}

\usepackage{natbib}
 
\newcommand{\boldtheta}{\boldsymbol \theta}

\begin{document}

\title{\LARGE Deep Reinforcement Learning-Based Precoding for Multi-RIS-Aided Multiuser Downlink Systems with Practical Phase Shift}

\author{Po-Heng Chou,~\IEEEmembership{Member,~IEEE}, Bo-Ren Zheng, Wan-Jen Huang,~\IEEEmembership{Member,~IEEE}, \\Walid Saad,~\IEEEmembership{Fellow,~IEEE}, Yu Tsao,~\IEEEmembership{Senior Member,~IEEE}, and Ronald Y. Chang,~\IEEEmembership{Senior Member,~IEEE}\vspace{-0.35in}
\thanks{This work was supported in part by the Academia Sinica under Grant 235g Postdoctoral Scholar Program, in part by the National Science and Technology Council (NSTC) of Taiwan under Grants 113-2218-E-110-008, 113-2218-E-110-009 and 113-2926-I-001-502-G, and in part by the U.S. National Science Foundation under Grant CNS-2030215. (Corresponding author: \emph{Wan-Jen Huang}).}

\thanks{Po-Heng Chou, Yu Tsao, and Ronald Y. Chang are with the Research Center for Information Technology Innovation (CITI), Academia Sinica (AS), Taipei, 11529, Taiwan (e-mail: d00942015@ntu.edu.tw, yu.tsao@citi.sinica.edu.tw, rchang@citi.sinica.edu.tw).}
\thanks{Bo-Ren Zheng and Wan-Jen Huang are with the Institute of Communications Engineering, National Sun Yat-sen University, Taiwan (e-mail: qazxc61109@gmail.com, wjhuang@faculty.nsysu.edu.tw).}
\thanks{Walid Saad is with the Bradley Department of Electrical and Computer Engineering, Virginia Tech (VT), VA, USA (e-mail: walids@vt.edu).}
}

\markboth{IEEE Wireless Communications Letters,~Vol.~14, No.~1, Jan.~2024}
{}

\maketitle
\begin{abstract}
This study considers multiple reconfigurable intelligent surfaces (RISs)-aided multiuser downlink systems with the goal of optimizing the transmitter precoding and RIS phase shift matrix to maximize spectrum efficiency.
A practical coupling effect is considered between the reflecting amplitude and phase shift for the reflecting elements and their corresponding reflectivities. 
However, with coupled optimization variables, the formulated optimization problem is non-convex.
Therefore, deep deterministic policy gradient (DDPG) based deep reinforcement learning (DRL) is proposed to solve the optimization problem. A practical scenario was simulated, whereby the proposed DDPG models were trained considering the random and fixed number of users. The simulation results show that the proposed DDPG, despite being comparably complex, outperforms the two optimization-based algorithms.
\end{abstract}
\vspace{-0.05in}

\IEEEpeerreviewmaketitle
\vspace{-0.2in}
\section{Introduction}

Millimeter-wave (mmWave)~\cite{Haider2022} frequency bands (i.e., 24--52 GHz) are integral components of wireless beyond 5G and 6G cellular systems. However, such high-frequency transmission bands are often characterized by high power consumption and signal energy attenuation caused by environmental obstacles because of their short wavelengths. 
Reconfigurable intelligent surfaces (RISs)~\cite{Omar2023, Peng2023, Yuqian2022, Feng2020} have been proposed to reflect transmitted signals from a base station (BS) to bypass obstacles and improve spectral and energy efficiency.
An RIS comprises several reflective elements with each reflective element having a different response because of phase changes. The practical reflectivity of the RIS has a definite impact on the system. Therefore, selecting the appropriate reflection phase is a critical challenge~\cite{Abeywickrama2020TWC}. Most previous works~\cite{Omar2023, Peng2023, Yuqian2022, Feng2020} have assumed that reflectivity equals one; however, this is not a practical assumption. Therefore, the relationship between the reflectivity and phase shift must be considered when designing an RIS controller, which significantly increases the complexity of the optimization.

An important challenge is the design and optimization of the precoding and RIS phase shift matrices in a multi-RIS-aided multiuser multiple-input single-output (MISO) system.
This is a non-convex optimization problem that cannot be solved effectively with techniques such as alternating optimization (AO) because these techniques have high computational complexity that does not scale properly with the number of optimization parameters (e.g., number of antennas, number of RISs, and reflective elements).
This results in a significant increase in computational time, and the results obtained are outdated solutions for varying wireless mmWave channels~\cite{Gong2022}. 
A complexity analysis of the AO for the RIS precoding design is shown in~\cite{Ruijin2023} for a practical phase shift model.
In~\cite{Cheonyong2024, Cheonyong2022}, the authors analyzed the achievable rate asymptotically and demonstrated the significant complexity of multi-RIS-aided multiuser systems for mmWave channels.
In addition to the optimization algorithm, reinforcement learning (RL)~\cite{Kasgari2021} can be adapted to dynamic environments by an agent exploring uncharted territories and utilizing existing knowledge~\cite{Kaelbling1996}. To adapt to varying wireless mmWave channels, the RIS controller employs RL~\cite{Omar2023, Peng2023, Yuqian2022, Feng2020}, and the
deep Q network (DQN)~\cite {Mnih2015} is a basic deep RL that has been widely adopted for the RIS precoding design~\cite {Peng2023}. However, the performance of DQN is poor because of the discrete action spaces, and thus, it cannot be trained well, especially when the number of users is random.
Therefore, we adopted a deep deterministic policy gradient (DDPG) for the scenario of the random number of users.
DDPG is a model-free, off-policy DRL proposed in~\cite{Silver2014}, combining the advantages of DQN and DPG-based actor critic~\cite{Konda1999}.
By operating over continuous action spaces simultaneously, it approximates the awareness of the full state space~\cite{Timothy2015}.
In studies related to DDPG-based RIS precoding, only a single RIS with an \emph{ideal} RIS reflective model has been considered in single-user~\cite{Yuqian2022} and multiuser~\cite{Feng2020} systems.
However, there are few studies addressing multi-RIS-aided multi-user MISO systems that simultaneously consider the \emph{practical} RIS reflective model~\cite{Abeywickrama2020TWC}.

The main contributions of this letter are as follows:
(a) Unlike~\cite{Yuqian2022} and~\cite{Feng2020}, we adopted DDPG to design the precoders of the RIS phase shift and transmitter for a multi-RIS-aided multiuser MISO system under a practical RIS reflective model.
(b) In comparisons with DDPG, a double DQN (DDQN)~\cite{Peng2023} and two optimization-based algorithms, namely, block coordinate descent (BCD)~\cite{BCD} and weighted minimum mean square error with power iteration (WMMSE-PI)~\cite{WMMSE-PI}, were implemented.
(c) The pre-trained scenarios of a fixed and random number of users, and the proposed DDPG trained with a random number of user equipment (UE) outperformed those trained with a fixed number of UE.

\vspace{-0.1in}
\section{System Model}
\begin{figure}
\centering
{\includegraphics[width=0.32\textwidth]{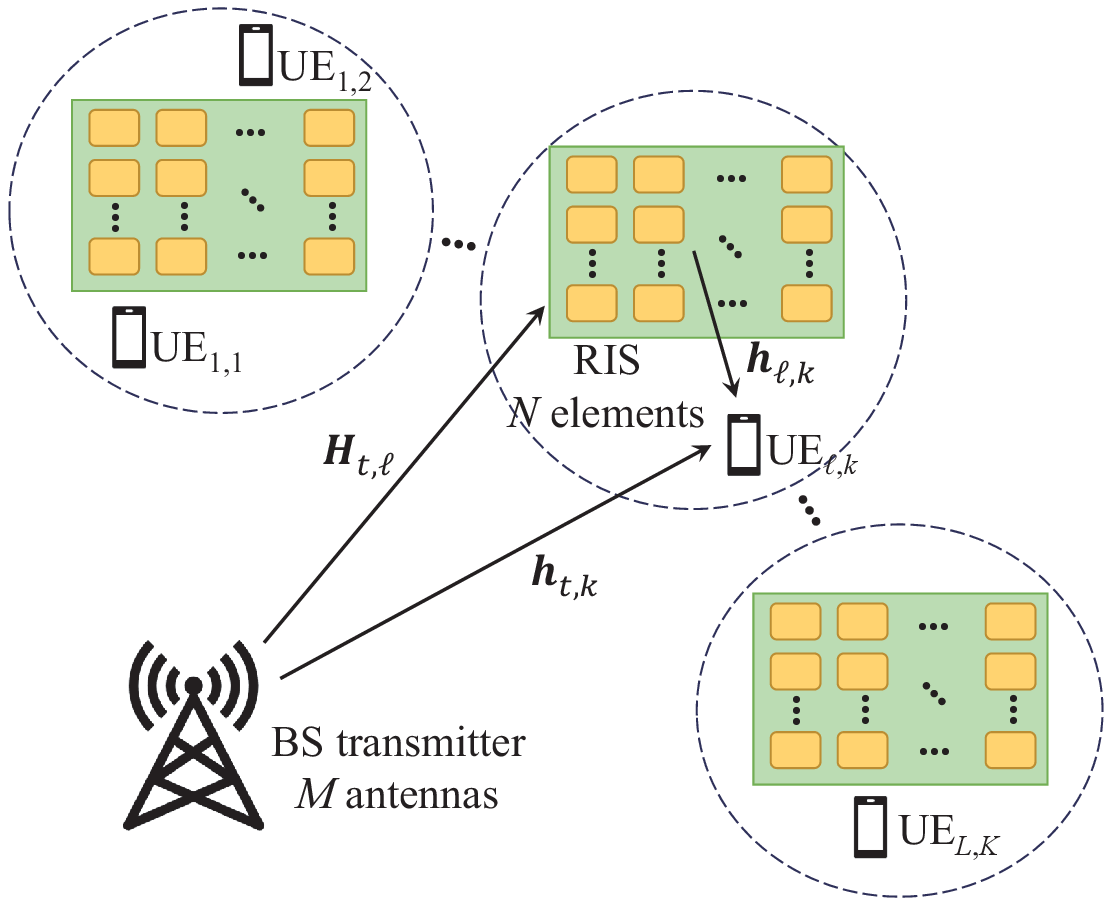}
\caption{The multi-RIS-aided multi-user downlink system.}\label{System_Model}}
\vspace{-0.15in}
\end{figure}

For the system model, we considered a multi-RIS-aided downlink system comprising a BS and several pieces of UE served by $L$ RISs distributed in the BS coverage area, as shown in Fig.~\ref{System_Model}. Each UE is served by one of the closest RISs.
The number of UEs served by RIS $\ell$ is $g_{\ell}$, and we define a set $\mathcal{G} = \{g_{1}, g_{2}, \ldots ,g_{L}\}$ for the number of UEs served by all RISs. The number of all UEs served by the BS is $K\triangleq\sum_{\ell = 1}^{L} g_{\ell}$.
The BS is equipped with $M$ antennas, and each UE has a single antenna.
Each RIS is equipped with $N=N_{x}\times N_{y}$ 
passive reflective elements, where $N_x$ and $N_y$ are the number of horizontal rows and vertical columns in the RIS array, respectively.

The signal received at UE $k$ can be expressed as:
\vspace{-0.1in}
\begin{equation}\label{equ:rx_sig}
y_{k} = \boldsymbol{h}_{k}\boldsymbol{w}_{k}x_{k} + \sum_{i = 1, i \neq k}^{K}\boldsymbol{h}_{k}\boldsymbol{w}_{i}x_{i} + n_{k},\vspace{-0.1in}
\end{equation}
where $x_k$ is the data symbol of UE $k$ with unit energy; $\boldsymbol{w}_{k}$ is the precoding vector for $x_k$; ${\boldsymbol h}_k$ is the effective channel coefficient from the BS to UE $k$; and $n_k\sim {\cal CN}(0, \sigma^{2})$ is the additive white noise at UE $k$. In Eq. (\ref{equ:rx_sig}), the first term is the desired signal for UE $k$, and the second term includes the inter-user interference. 
The effective channel coefficient $\boldsymbol{h}_{k}$ is given by
$\boldsymbol{h}_{k} = \boldsymbol{h}_{t,k}+ \boldsymbol{h}_{r,k}$
where $\boldsymbol{h}_{t,k}$ is the channel coefficient of the direct path between BS and UE $k$, and $\boldsymbol{h}_{r,k}$ is the composite channel from the BS to UE $k$ with reflections from the RISs. 
The channel coefficient of the direct path followed the mmWave Saleh-Valenzuela (SV) channel model, which is given by 
\begin{equation}
\boldsymbol{h}_{t,k} = \sqrt{\frac{PL_{t,k}}{N_{t,k}L_{t,k}}}\sum_{i=1}^{N_{t,k}}\sum_{q=1}^{L_{t,k}} \breve{\alpha}_{i,q}{\bf a}_{M}^H(\breve{\theta}_{i,q}^{(D)}),
\end{equation}
where  $PL_{t,k}$ is the path loss factor; $N_{t,k}$ and $L_{t,k}$ are the number of non-line-of-sight (NLOS) clusters and the number of paths in each cluster of the direct path from the BS to the UE $k$, respectively; $\breve{\alpha}_{i,q}\!\sim\!{\cal CN}(0,1)$ is the gain of the $q$-th path in the $i$-th cluster. The vector ${\bf a}_{M}(\theta)=[1, e^{j\frac{2\pi d}{\lambda}\sin\theta}, \cdots, e^{j\frac{2\pi d}{\lambda}(M-1)\sin\theta}]^T$ is the response vector for a uniform linear array, and $\breve{\theta}_{i,q}^{(D)}$ is the angle of departure (AOD) at the BS, where $d$ is the antenna spacing at the BS and $\lambda$ denotes the carrier wavelength.
The composite channel from the BS to UE $k$ with reflections from multiple RISs is  expressed as
\vspace{-0.1in}
\begin{align}\label{equ:composite_channel}
\boldsymbol{h}_{r,k} = \sum_{\ell=1}^{L}\boldsymbol{h}_{\ell ,k}\boldsymbol{\Theta}_{\ell}\boldsymbol{H}_{t,\ell},\vspace{-0.1in}
\end{align}
where $\boldsymbol{H}_{t,\ell} \in \mathbb{C}^{N \times M}$ and $\boldsymbol{h}_{\ell ,k} \in \mathbb{C}^{1 \times N}$
are the channels from BS to RIS $\ell$ and from RIS $\ell$ to UE $k$, respectively.
The SV channel matrices are as follows:
\begin{align}
\boldsymbol{H}_{t,\ell} &\!=\! \sqrt{\frac{PL_{t,\ell}G_{\rm RIS,\ell}}{N_{t,\ell}L_{t,\ell}}}\sum_{i=0}^{N_{t,\ell}}\sum_{q=1}^{L_{t,\ell}} \alpha_{i,q} {\bf a}_{N_x,N_y}(\phi_{i,q}^{(A)},\theta_{i,q}^{(A)}) {\bf a}_{M}^H(\theta_{i,q}^{(D)}),\nonumber\\
\boldsymbol{h}_{\ell ,k} &\!=\!\sqrt{\frac{PL_{\ell,k}G_{\rm RIS,\ell}}{N_{\ell,k}L_{\ell,k}}}\sum_{i=0}^{N_{\ell,k}}\sum_{q=1}^{L_{\ell,k}} \tilde{\alpha}_{i,q} {\bf a}^H_{N_x,N_y}(\phi_{i,q}^{(D)},\theta_{i,q}^{(D)}),\label{equ:SV}
\end{align}
where $PL_{t,\ell}$ and $PL_{\ell,k}$ are the loss factors of the links from BS to RIS $\ell$ and from RIS $\ell$ to UE $k$; $G_{\rm RIS,\ell}$ is the gain compensation at RIS $\ell$; $(N_{t,\ell}, L_{t,\ell})$ and $(N_{\ell,k},L_{\ell,k})$ are the number of clusters and the number of paths in each cluster, respectively; $\alpha_{i,q}\sim{\cal CN}(0,1)$ and $\tilde{\alpha}_{i,q}\sim{\cal CN}(0,1)$ are the gains of the $q$-th path in the $i$-th cluster of the two links. 
The zeroth cluster in Eq. (\ref{equ:SV}) is assumed to refer to the line-of-sight (LOS) paths of each link. 
The vector  ${\bf a}_{M,N}(\phi,\theta)=[1,\cdots$, $e^{j\frac{2\pi d}{\lambda}(p\sin\phi\sin\theta+q\cos\theta)}$, $\cdots$, 
 $e^{j\frac{2\pi d}{\lambda}((M-1)\sin\phi\sin\theta+(N-1)\cos\theta)}]^T$
is the response vector of a uniform planar array; $\theta_{i,q}^{(D)}$ is the AOD at the BS; $(\phi_{i,q}^{(A)},\theta_{i,q}^{(A)})$ are the azimuth and elevation angles of arrival (AOA) at the RIS and $(\phi_{i,q}^{(D)},\theta_{i,q}^{(D)})$ are the azimuth and elevation AoDs at the RIS, respectively.


%

In Eq. (\ref{equ:composite_channel}), $\boldsymbol{\Theta}_{\ell} = \textrm{diag} (\beta_{\ell ,1}e^{j\theta_{\ell ,1}}, \beta_{\ell ,2}e^{j\theta_{\ell ,2}}, \ldots , \beta_{\ell ,N}e^{j\theta_{\ell ,N}} ) \in \mathbb{C}^{N \times N}$ is the reflection response of RIS $\ell$, where $\textrm{diag} (\cdot)$ denotes the diagonal operation of a matrix; $\theta_{\ell , n} \in [- \pi, \pi]$ and $\beta_{\ell ,n} \in (0, 1]$ are the phase shift and amplitude response of the $n$-th reflective component in the $\ell$-th RIS, respectively.
The literature~\cite{Omar2023, Peng2023, Yuqian2022, Feng2020}, mostly assumes that the amplitude response is ideal, i.e., $\beta_{\ell,n}=1$.
In contrast, we considered a practical relationship between the amplitude response and phase shift, as proposed in \cite{Abeywickrama2020TWC}. Specifically,
the amplitude response $\beta_{\ell, n}$ is a function of the phase shift $\theta_{\ell, n}$, which can be approximated by 
\begin{align}
\beta_{\ell, n} = (1-\beta_{\min})\left(\frac{\sin(\theta_{\ell, n}-\phi_{0})+1}{2}\right)^{\alpha}+\beta_{\min},
\label{Practical_Model}
\end{align}
for $\theta_{\ell, n}\in [\theta_{\min}, \theta_{\max}]$, where $\alpha$ is a parameter that controls the steepness of the function curve; $\beta_{\min}$ is the minimum amplitude; $\phi_{0}$ is the difference
between the phase with minimum amplitude and $-\pi$; $\theta_{\min}$ and $\theta_{\max}$ are the lower and upper bound of the phase shift imposed by hardware limitations. When $\alpha = 0$, we have $\beta_{\ell, i} = 1$, which reduces to the ideal case of reflective amplitude.

\vspace{-0.1in}
\section{Problem Formulation}

Our goal was to jointly optimize the precoding vectors $\boldsymbol{w}_{k}$ at the BS and the phase shifting vector of the RISs $\boldsymbol{\Theta}_{\ell}$, such that the sum rate of the multi-RIS-aided multi-user downlink system is maximized.
The achievable rate of UE $k$ is given by:
\begin{align}
{C_k}=\log_{2}\left(1+\frac{|\boldsymbol{h}_{k}\boldsymbol{w}_{k}|^{2}}{\sum_{i = 1, i \neq k}^{K}|\boldsymbol{h}_{k}\boldsymbol{w}_{i}|^{2} + \sigma^{2}}\right) ~~\rm bps/Hz.
\end{align}
This joint optimization can be formalized as follows:
\begin{subequations}
\label{optimization_problem_all}
\begin{align}
&\max_{\boldsymbol{w}_k, \boldsymbol{\Theta}_{\ell}} \;\; \sum_{k=1}^{K} C_k,\\
&\textrm{s.t}\;\; \textrm{tr}(\boldsymbol{W}\boldsymbol{W}^{H})\leq P_{\max}\\
&\theta_{\ell, n} \in [\theta_{\min}, \theta_{\max}]
\label{optimization_problem}
\end{align}
\end{subequations}
where ${\boldsymbol{W}}=[{\boldsymbol{w}}_1,{\boldsymbol{w}}_2, \cdots,{\boldsymbol{w}}_K]$, $\textrm{tr}(\cdot)$ is the trace of a matrix, and 
$P_{\max}$ is the maximal transmission power at the BS.
The optimization problem in (\ref{optimization_problem_all}) is non-convex. To solve this problem, AOs can be adopted to iteratively find a sub-optimal solution. However, the iterative operation of the AO imposes a significant computational burden and leads to additional latency. To speed up the optimization procedure, the DDPG method was adopted to jointly optimize the precoding matrix $\boldsymbol{W}$ and phase shifting parameters of RISs $\boldsymbol{\Theta}_{\ell}$. This is facilitated by the fact that the trained RL model requires only matrix operations.
\vspace{-0.1in}
\section{Deep Deterministic Policy Gradient-Based Joint Optimization}
The DDPG~\cite{Timothy2015} comprises a policy and a target network, where each network has two sub-networks: an actor and a critic sub-network. The policy network
explores the environment and stores the experience in the
replay buffer. The policy network is then updated by temporal-difference (TD) learning in the direction of the gradient. TD learning is used to compute the loss function during the training phase. The target network estimates the value (expected reward) using a possible future state. The details of the DDPG flow for optimizing the RIS-aided MISO system are shown in Fig.~\ref{DDPG}. The DDPG tasks are detailed below:
\begin{figure}[t]
\centering
{\includegraphics[width=0.4\textwidth]{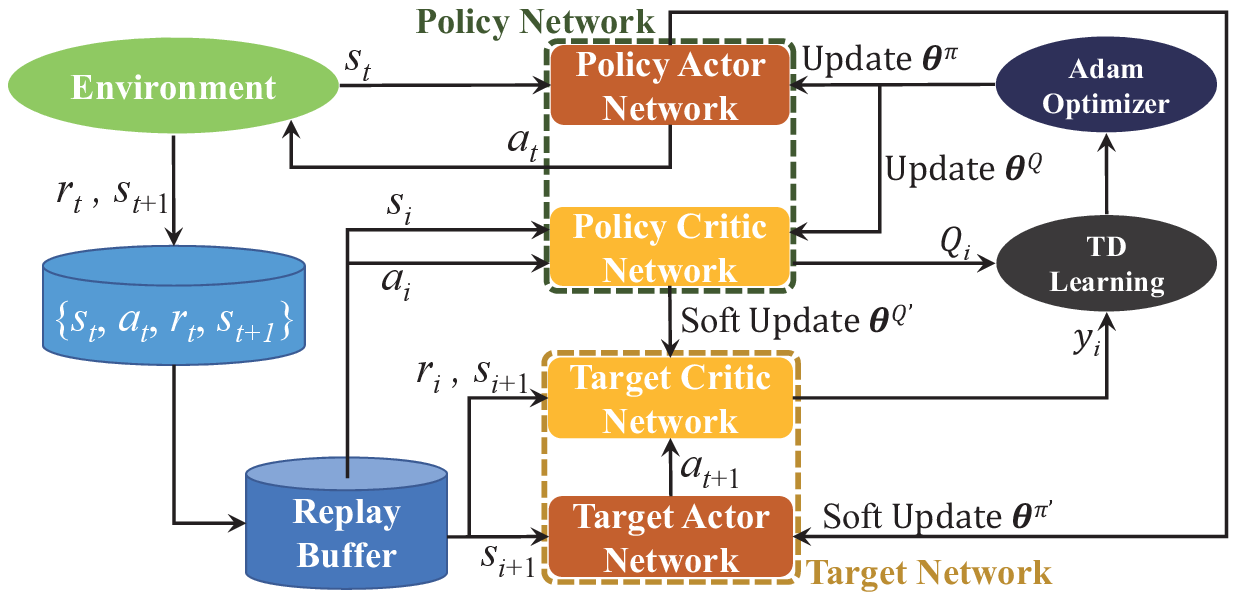}
\caption{The architecture of DDPG.}\label{DDPG}}
\vspace{-0.15in}
\end{figure}
\textbf{Data Preprocessing:}
Because neural networks only consider real values, we separated the input complex value data into real and imaginary parts.
The optimization parameters ($\boldsymbol{W}, \boldsymbol{\Theta}_{\ell}$) in action space can be defined as $a_{t} = \{{\rm Re}(\boldsymbol{w}_{1}^{T}, \ldots , \boldsymbol{w}_{K}^{T}), {\rm Im}(\boldsymbol{w}_{1}^{T}, \ldots , \boldsymbol{w}_{K}^{T}),$ ${\rm Re}(\theta_{1,1}, \ldots , \theta_{L,N}), {\rm Im}(\theta_{1,1}, \ldots , \theta_{L,N})\}$, where the size of the action space is $2(MK+LN)$.
Thus, we can calculate $\boldsymbol{w}_{k} \leftarrow \boldsymbol{u}_{k}\boldsymbol{w}_{k}, k = 1, \ldots , K$.
To satisfy the constraints~(\ref{optimization_problem_all}b) and~(\ref{optimization_problem_all}c), the variables $\boldsymbol{W} \leftarrow \sqrt{P_{\max}}\frac{\boldsymbol{W}}{\|\boldsymbol{W}\|_{F}}$ and $\theta_{\ell, n} \leftarrow \frac{\theta_{\ell, n}}{|\theta_{\ell, n}|}$ are normalized for the outputs. 
Considering a practical RIS reflective model, $\theta_{\ell, n}^{(p)}$ needs to be further merged with the amplitude response using
\begin{align}
\theta_{\ell, n}^{(p)}\leftarrow \left\{
\begin{array}{ll}
\beta_{\ell,n}\theta_{\ell,n}, &\theta_{\min} \leq \theta_{\ell,n} \leq \theta_{\max},\\
\beta_{\ell,n}\theta_{\max}, &\theta_{\ell,n} > \theta_{\max},\\
\beta_{\ell,n}\theta_{\min}, &\theta_{\ell,n} < \theta_{\min},
\end{array}\right.
\end{align}
where $\beta_{\ell,n}$ is given by~(\ref{Practical_Model}). In case of the ideal RIS reflective model, $\theta_{\ell, n}^{(i)}=\theta_{\ell, n}$.

Channel state information (CSI) is typically used as state input data.
However, the large state space of multi-user systems consumes considerable computational power. 
Because this action is strongly related to the information in the UEs and the resultant achievable rate, 
 the policy and target critic networks must refer to the previous action to determine the current state. 
Thus, we define $s_{t} = \{\mathcal{U}, a_{t-1}, \mathcal{R}_{t-1}\}$ as a state, where $\mathcal{U} = \{u_{1}, \ldots, u_{K}\}$ is the UE information, with $u_k\in\{0,1\}$ denoting the presence of the $k$-th UE for previous action $a_{t-1}$ and achievable rates of previous actions $\mathcal{R}_{t-1} = \{r_{t-1,1}, \ldots,r_{t-1, K}\}$, where $r_{t-1,k}$ is the previous reward of the $k$-th UE after the previous action.
Compared to the state with full CSI, the dimensionality of the state space is reduced from $2(MKN+LN+K)$ to $2(MK+LN+K)$.
The objective is to maximize the achievable rate, which is defined as the reward $r_{t} = \sum_{k=1}^{K}C_k$.
\textbf{Environmental Exploration:} The current state $s_{t}$ from the environment is input into the policy actor sub-network, which outputs the current action $a_{t} = \pi (s_{t}|\boldtheta^{\pi})$, where $\pi$ is the deterministic policy of action selection, and $\boldtheta^{\pi}$ denote the weights of the policy actor sub-network.
To explore diversity, we added noise $n_{t}\thicksim \mathcal{N}(0, 0.1)$ to $a_{t} \leftarrow a_{t} + n_{t}$.
By performing the current action $a_{t}$, the current reward $r_{t}$ and next state $s_{t+1}$ are obtained from the environment, thereby storing $\{s_{t}, a_{t}, r_{t}, s_{t+1}\}$ in the replay buffer.
\textbf{Experience Learning:} Based on the size of the mini-batch, the agent randomly samples $D$ data points for training.
Policy critic sub-network estimates the current Q value $Q_{i} = Q(s_{i},a_{i}|\boldtheta^{Q})$, where $\boldtheta^{Q}$ are the weights of the policy critic sub-network, and $i = 1, \ldots , D$.
The next stage is input to the target actor sub-network, which outputs the next action $a_{i+1} = \pi (s_{i+1}|\boldtheta^{\pi})$.
The target critic sub-network estimates the next Q value $Q_{i+1}' = Q'(s_{i+1}, \pi(s_{i+1}|\boldtheta^{\pi'})|\boldtheta^{Q'})$, where $\boldtheta^{\pi'}$ and $\boldtheta^{Q'}$ are the weights of the target actor and critic sub-networks, respectively. The training target Q value is defined as
\begin{align}
y_{i} = \left\{
\begin{array}{ll}
r_{i} + \gamma Q_{i+1}'(s_{i+1}, \pi'(s_{i+1}|\boldtheta^{\pi'}) |\boldtheta^{Q'}), &i = 1, \ldots,  D-1\\
r_{i}, &i = D,
\end{array}\right.
\label{equ:q-value}
\end{align}
where $\gamma$ is the discount factor.
\textbf{Weight Update:}
To minimize the loss function, the policy critic sub-network updates $\boldtheta^{Q}$ by TD learning. The loss function is defined as follows:
\begin{align}
Loss = \frac{1}{D}\sum_{i = 1}^{D}(Q_{i}(s_{i},a_{i}|\boldtheta^{Q}) - y_{i})^{2}.
\label{equ:loss}
\end{align}
To maximize the Q value, the policy actor sub-network updates $\boldtheta^{\pi}$ using the policy gradient:
\begin{align}
\nabla_{\boldtheta^{\pi}}\mathcal{J} = \frac{1}{D}\sum_{i = 1}^{D}\nabla_{a_{i}}Q_{i}(s_{i},a_{i}|\boldtheta^{Q})\nabla_{\boldtheta^{\pi}}\pi(s_{i}|\boldtheta^{\pi}),
\label{equ:grad}
\end{align}
where $\mathcal{J} = \mathbb{E}[r_{1}|\pi]$ is the cumulative discounted reward accumulated after the start state.

Finally, the weights of the target network ($\boldtheta^{\pi '}$ and $\boldtheta^{Q'}$) are updated iteratively until they converge using the soft update as
\begin{align}
\left\{
\begin{array}{l}
\boldtheta^{\pi'} \leftarrow \tau\boldtheta^{\pi} + (1-\tau)\boldtheta^{\pi '}\\
\boldtheta^{Q'} \leftarrow \tau\boldtheta^{Q} + (1-\tau)\boldtheta^{Q'},
\end{array}\right.
\label{equ:soft}
\end{align}
where $\tau$ is the soft update factor that is used to control the target network learning rate.
The pseudo-code for the DRL-based framework is presented in \textbf{Algorithm 1}. Because the action space of DDPG is continuous, only a sub-optimal solution is obtained.

\begin{algorithm}
\footnotesize
\caption{The DRL-based Framework}\label{alg:DRL}
\SetKwInput{KwData}{Input}
\SetKwInput{KwResult}{Output}
\KwData{Discount factor $\gamma$, soft update factor $\tau$, mini-batch size $D$, learning rate $\alpha$, and replay buffer size $B$.\\
Randomly initialize the policy critic sub-network $Q(s,a|\boldtheta^{Q})$ and policy actor sub-network $\pi (s|\boldtheta^{\pi})$.\\
Initialize the target policy network $Q'(s,a|\boldtheta^{Q'})$ and target actor sub-network $\pi'(s|\boldtheta^{\pi'})$ with the corresponding policy networks.}
 \For{\rm episode $i = 1, \ldots , I$}{
 Obtain the current UE information $\mathcal{U} = \{u_{1}, \ldots , u_{K}\}$ (presence: 1 and absence: 0)\;
 Randomly select phase shifts to obtain $\boldsymbol{W}^{(0)}$ and $\boldsymbol{\Theta}_{\ell}^{(0)}$ as initial state $s_{1} = \{\mathcal{U}, a_{0}, \mathcal{R}_{0}\}$\;
 Generate a random noise $n_{t}\thicksim \mathcal{N}(0, 0.1)$\;
 \For{\rm time step $t = 1, \ldots , T$ }{
 Action $a_{t} \leftarrow \pi (s_{t+1}|\boldtheta^{\pi }) + n_{t}$.\\
 Perform $a_{t}$ into $\boldsymbol{\Theta}_{\ell}$ to obtain the achievable rates $\mathcal{R}_{t} = \{r_{t,1}, \ldots ,r_{t,K}\}$ and the next state $s_{t+1}$.\\
 Store the transition $\{s_{t}, a_{t}, r_{t}, s_{t+1}\}$ into $B$.\\
 Sample a $D$ minibatch transitions $\{s_{i}, a_{i}, r_{i}, s_{i+1}\}$ from $B$.\\
 Set target Q value based on~(\ref{equ:q-value}).\\
 Update $Q(s,a|\boldtheta^{Q})$ by minimizing the loss in~(\ref{equ:loss}).\\
 Update the policy $\pi (s|\boldtheta^{\pi})$ using the sampled policy gradient in~(\ref{equ:grad}).\\
 Soft update the target networks based on~(\ref{equ:soft}).\\
 Update the sate $s_t = s_{t+1}$. 
 }
 }
\KwResult{The sub-optimal solutions of precoding matrix $\boldsymbol{W}^{*}$ and RIS phase shift matrix $\boldsymbol{\Theta}_{\ell}^{*}$.}
\end{algorithm}

\section{Simulation Results and Analysis}
The following setup was considered in the simulations:
At the BS, the number of antennas was 16 and the transmission power was 20 dBm.
We deployed four RISs that were uniformly distributed on the circumference of a quarter circle with a radius of 100 m centered at the BS.
The details of the LOS and NLOS path loss parameters for the mmWave channel can be defined by referring to 3GPP standard release 16~\cite{Release16}.
The distance between the RISs was 39.27 meters.
Each RIS had 16 reflective elements with a uniform distribution of $U\thicksim [1, 4]$ UEs, which appeared randomly around the RIS.
For the AOAs and AODs of all propagation paths, the parameter settings in \cite{SV_Channel} were adopted.
To reduce the uncertainty in learning, the distances between the RIS and BS/UE were fixed at 100 m and 2 m, respectively.
The hyperparameters were set as follows: $I = 1000$, $T = 2000$, discount coefficient $\gamma = 0.95$, soft update factor $\tau = 0.0005$, mini batch size $D = 128$, and learning rate $\alpha = 0.0001$.
After an episode of training, DDPG performed testing and recorded the reward obtained from a total of 1,000 environments (all combinations of $U\thicksim [4,16]$ UEs in the four RISs).
The weights corresponding to the current highest reward were recorded and updated per episode until training was completed.
The actor and critic sub-networks in the policy network were structurally identical to those in the target network. {\bf The actor sub-network} consisted of four layers of neural networks: an input layer, two fully connected layers with 1024 neural elements each, and an output layer. For faster convergence, in addition to the input layer, the output of each layer was connected to a normalization layer and an activation function. The fully connected layers utilized the ReLU function, and the output layer employed the tanh function.
{\bf The critic sub-network} included five layers of neural networks incorporating both state and action input layers. This sub-network was organized as follows. The input layer was split into state and action layers, and their outputs were combined through an additional layer. The output of this additional layer was connected to a normalization layer and an activation function. The output layer generated the Q value through a fully connected layer that connected the normalization layer and activation function.



\begin{figure}
\vspace{-0.15in}
\centering
{\includegraphics[width=0.45\textwidth]{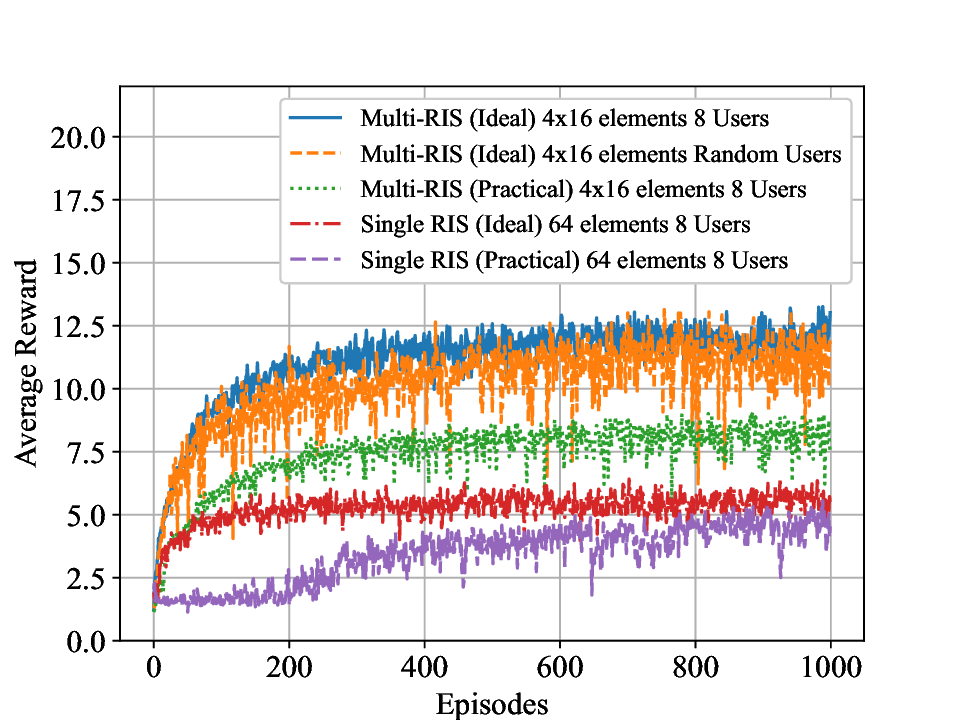}
\caption{Training rates of the proposed DDPG for single- and multi-RIS aided under practical and ideal reflective considerations.}
\label{Figure_4-2}}	
\vspace{-0.1in}
\end{figure}

Fig.~\ref{Figure_4-2} shows the convergence of the DDPG algorithm in single- and multi-RIS-aided downlink systems based on practical and ideal reflective considerations. Specifically, the single RIS system comprises 64 elements and serves eight UEs.
The multi-RIS system involves four RISs, each consisting of 16 elements, and serves two UEs. 
The training rewards tended to converge after 400 episodes under the ideal and practical reflective models.
Due to amplitude attenuation and the limited range of phase shifts, the achievable rate is degraded in the case of a practical reflective model.

\begin{figure}
\centering
{\includegraphics[width=0.4\textwidth]{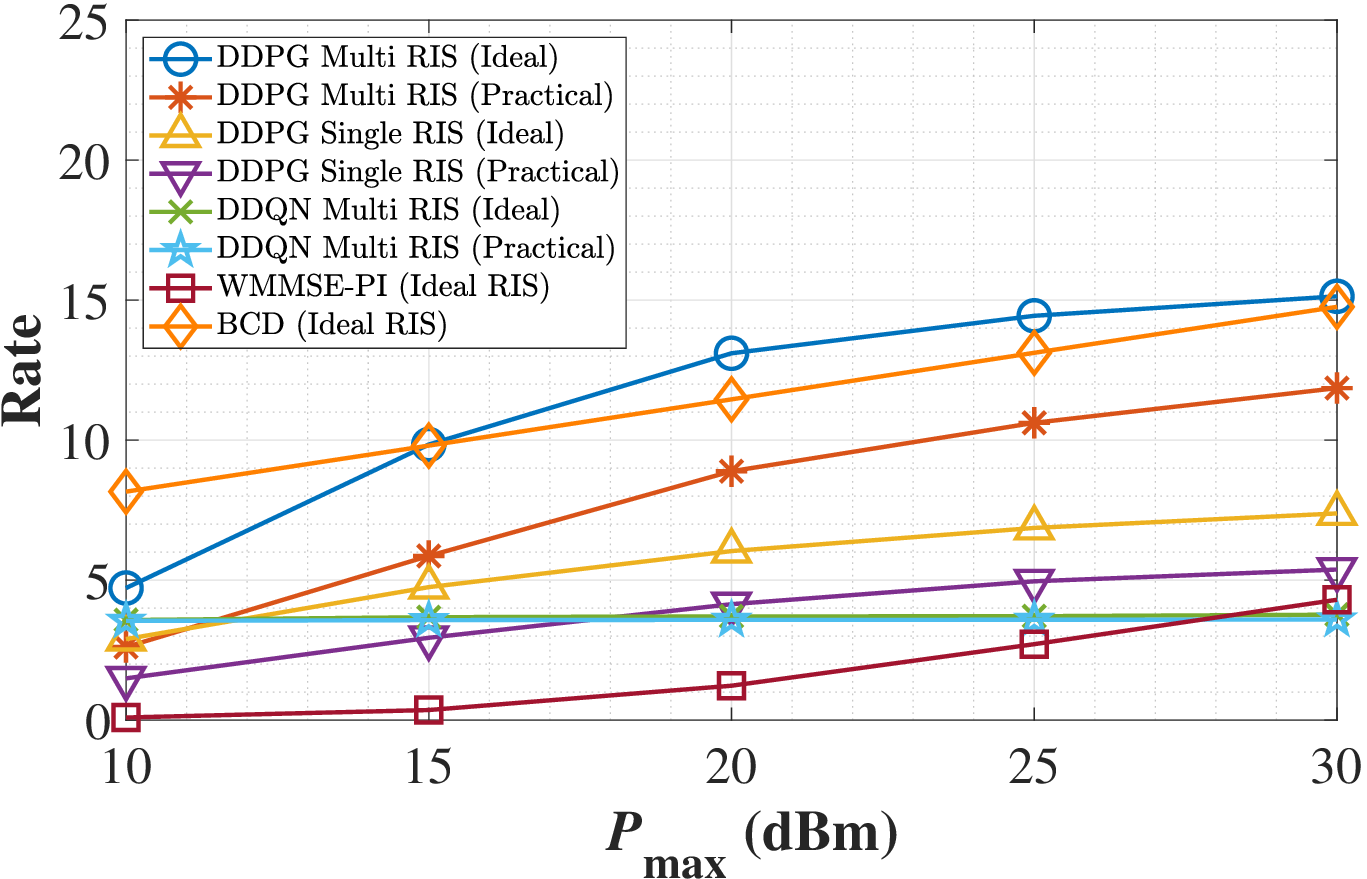}
\caption{Comparison of the proposed DDPG, BCD, and WMMSE-PI.}\label{Figure_4-3}}
\vspace{-0.1in}
\end{figure}

Fig.~\ref{Figure_4-3} compares the achievable rates of the proposed DDPG with those of two optimization-based algorithms, WMMSE-PI~\cite{WMMSE-PI}, and BCD~\cite{BCD}, and DDQN~\cite{Peng2023}, by averaging over 1,000 channel realizations.
In particular, the proposed DDPG was applied to single- and multi-RIS-aided downlink systems under either ideal or practical reflective models, and the sizes of the RISs were identical to those shown in Fig.~\ref{Figure_4-2}.
Notably, WMMSE-PI~\cite{WMMSE-PI}, BCD~\cite{BCD}, and DDQN~\cite{Peng2023}  algorithms are only applicable to cases with a single RIS and fixed UE under the ideal reflective assumption.

During the online inference phase, the proposed DDPG requires $3.71 \times 10^6$ floating-point operations (FLOPs) on average. 
For a fair comparison, we compared BCD, WMMSE-PI, DDQN, and DDPG with full convergence.
Fig.~\ref{Figure_4-3} shows that the achievable rate of the case with practical RISs is lower than the case with ideal RISs because of the substantial amplitude attenuation of the reflective signals. However, the case with multiple RISs outperforms the case with a single RIS because additional spatial diversity can be exploited and inter-user interference can be further suppressed. In the following comparison, we selected $P_{\max} = 20$ dBm.
The achievable rate of the multi-RIS outperformed the single-RIS by $117\%$ and $114\%$ for the ideal and practical RISs, respectively.
For the multi-RIS system under the ideal phase shift assumption, DDPG outperformed DDQN by $63\%$, showing a disparity between discrete and continuous action spaces.
For the single-RIS system under the ideal phase shift assumption, the achievable rate of DDPG outperformed that of the WMMSE-PI by $394.3\%$.
In addition, the achievable rate of BCD outperformed DDPG by $89.9\%$.
However, the computational complexity of BCD ($7.03 \times 10^8$ FLOPs) is much higher than that of the testing phase of DDPG.



\begin{figure}[t]
\centering
{\includegraphics[width=0.4\textwidth]{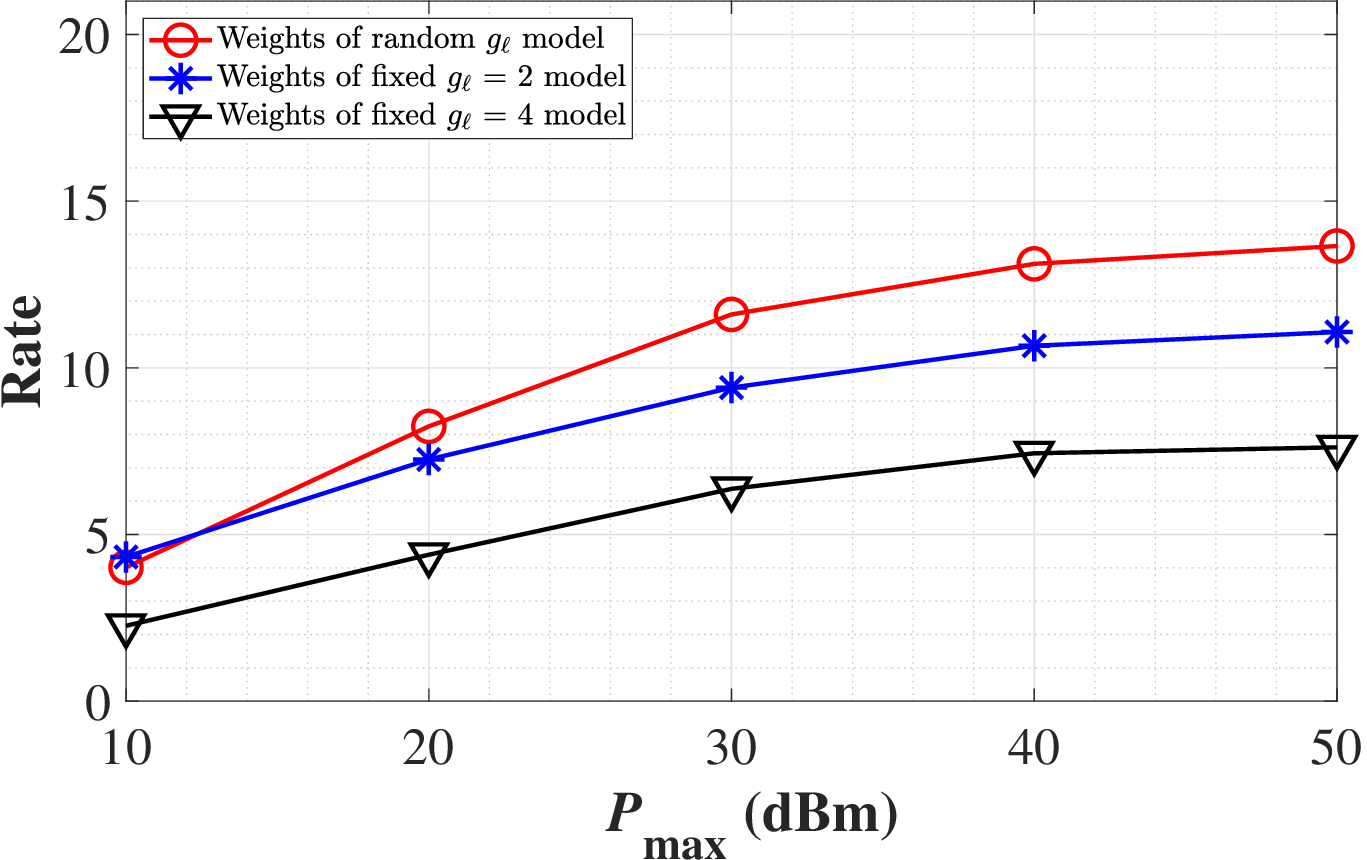}
\caption{Comparison of random and fixed UE amounts in the training phases.}\label{Figure_4-5}}	
\vspace{-0.3in}
\end{figure}

Fig.~\ref{Figure_4-5} shows that the number of UEs is random or fixed ($g_{\ell} = 2$ or $4$) during the training phase. The trained weights used from these three scenarios were applicable to the scenario of the random UE number $U\thicksim[4, 16]$ and different maximum transmission power constraints in the testing phase. Comparing the trained weights from scenarios with different numbers of UEs, the proposed DDPG model trained with a random number of UEs outperformed that trained with a fixed number of UEs. When $P_{\max} = 20$ dBm, the performance of the training weights with a random number of UEs outperformed that of the training weights with a fixed number of UEs for $g_{\ell} = 2$ and $4$ by $23.3\%$ and $82.1\%$, respectively. When the training is based on a specific user number, the DDPG model lacks sufficient data from other scenarios, resulting in degraded performance. In addition, the difficulty of optimization in multi-user systems is related to the amount of interference among the UEs.

\vspace{-0.1in}
\section{Conclusion}
In this letter, we presented a multi-RIS-aided downlink system with uniformly distributed randomly positioned UEs. To address practical considerations, the limitations in the phase shift of RIS reflective elements were considered by adjusting the coefficients in the approximation equation.
The DDPG algorithm was employed to optimize both active and passive precoding. The simulation results demonstrate that the proposed DDPG model can tolerate slight variations in the transmission power, channel conditions, and UE mobility.
In future work, we will extend the DDPG algorithm to a simultaneously transmitting and reflecting (STAR)-RIS-aided MIMO system~\cite{Omar2023} with UE mobility consideration.


\ifCLASSOPTIONcaptionsoff
 \newpage
\fi

\bibliographystyle{IEEEtran}





\end{document}